# Absence of stable atomic structure in fluorinated graphene

Danil W. Boukhvalov[1,2]

[1]*Department of Chemistry, Hanyang University, 17 Haengdang-dong, Seongdong-gu, Seoul 133-791, Korea*
[2]*Theoretical Physics and Applied Mathematics Department, Ural Federal University, Mira Street 19, 620002 Ekaterinburg, Russia*

*Based on the results of first-principles calculations we demonstrate that significant distortion of graphene sheets caused by adsorption of fluorine atoms leads to the formation of metastable patterns for which the next step of fluorination is considerably less energetically favorable. Existence of these stable patterns oriented along the armchair direction makes possible the synthesis of various $CF_x$ structures. The combination of strong distortion of the nonfluorinated graphene sheet with the doping caused by the polar nature of C–F bonds reduces the energy cost of migration and the energy of migration barriers, making possible the migration of fluorine atoms on the graphene surface as well as transformation of the shapes of fluorinated areas. The decreasing energy cost of migration with increasing fluorine content also leads to increasing numbers of single fluorine adatoms, which could be the source of magnetic moments.*

## 1. Introduction

Functionalization of graphene is used to tune its electronic, optical, mechanical and chemical properties [1]. The best known kind of functionalized graphene is graphene oxide, which can be obtained by means of exfoliating graphite oxide and was first synthesized more than 150 years ago [2]. Unfortunately, graphene oxide's complicated method of synthesis and inexact chemical structure make it unsuitable for some prospective applications; structural problems of particular note include the presence of perforations of various sizes and shapes, decreasing size of the graphene flakes and nonuniform distribution of functionalized areas [2]. The main causes of these nonuniformities are the possibility of divalent oxygen to form epoxy and hydroxyl groups on the surface of graphene, energetic unfavorability of the total oxidation of graphene [3], and formation of vacancies in the graphene membrane by means of removal of C atoms as CO and $CO_2$ gas molecules [4]. For these reasons, the use of a monovalent functionalizing species seems more attractive for uniform and controllable functionalization of graphene. The simplest choice for this is hydrogen. Hydrogenation of graphene uniformly and on both sides to form graphene was predicted theoretically [5,6] and realized experimentally soon thereafter [7]. The technical difficulties in the hydrogenation of graphene are serious, including the requirement of using hydrogen plasma to obtain uniform coverage at high concentrations; these difficulties have necessitated the use of a substitute for hydrogen. Aryl species could provide rather uniform functionalization of graphene with the opening of narrow band gap [8]. Weak points of aryl functionalization include the limited level of functionalization owing to the larger size of the adsorbed species and van der Waals repulsion between them [9], and the ability to functionalize only one side for technical reasons [8].

Fluorine seems to be a reasonable alternative functionalizing species, and its use has been discussed previously. Many previous reports on the fluorination of graphite [10-14] have also suggested the feasibility of using fluorine to chemically modify graphene. Recent experimental reports have demonstrated unusual properties of fluorinated graphene that make it attractive for various applications [13] and have included observations of tremendous differences between hydrogenated and fluorinated graphene. In contrast to the case of hydrogenation, whereby single adatoms, small clusters [15], or total hydrogenation has been obtained for free-standing graphene, fluorination provides formation of a variety of $CF_x$ structures for x ranging from 0.10 to 0.98 [10,16,17]. Obtaining 100% fluorination is rather difficult. X-ray measurements evidence the formation of linearly patterned fluorinated areas along the graphene axis in semi-fluorinated graphene, but with no preferred axis [18,19]. Further experiments have demonstrated that in semi-fluorinated graphene, about 20% of fluorine adatoms do not belong to any uniform fluorination pattern but instead form small clusters, pairs, and, for a few percent, single adatoms [20,21]. Another recent experimental results also demonstrate tendency to formation of various patterns in fluorinated graphene [22] and its instability. [23]

Another poorly understood aspect of fluorinated graphene is its magnetic properties. In contrast to nonmagnetic graphane, fluorinated graphene is usually paramagnetic and the magnetic centers increase in number with increasing fluorination extent [16]. This experimental result is in conflict with a previously developed model whereby adatoms are thought to passivate dangling bonds [6], causing decay in and eventual vanishing of magnetism with increasing adatom concentration.Recent theoretical works discussed only several marginal cases such as sinle fluorine adatom [24,25], some possible configuration of semiflurinated graphene [26-28] and the case of total fluorination [29-31].

Motivated by this unsolved problem of fluorinated graphene we have performed further explorations in the present work, listed as follows. (i) We have compared the results of step-by-step modeling of graphene fluorination and hydrogenation and herein discuss their differences in terms of lattice distortions. (ii) We calculate the energy costs of and barriers to migration of pairs and single atoms of fluorine for fluorographenes having various fluorination extents. (iii) We estimate the energetics of transformations of various patterns of fluorine on graphene. (iv) We examine interactions between the extent of fluorination and the formation energetics of magnetic configurations.

## 2. Computational model and method

Modeling was performed using density functional theory (DFT), implemented by means of the pseudopotential code SIESTA [32], as was done in our previous work [6,33]. All calculations were performed using the generalized gradient approximation (GGA-PBE) including spin polarization [34]. Full optimization of the atomic positions was carried out. During this optimization, the ion cores were described by norm-conserving nonrelativistic pseudopotentials [35] with cutoff radii 1.14, 1.45 and 1.25 a.u. for C, F, and H, respectively; the wavefunctions were expanded with a double-$\zeta$ plus polarization basis of localized orbitals for C and F, and with a double-$\zeta$ basis for H. Optimizations of the force and total energy were performed with the accuracies of 0.04 eV/Å and 1 meV, respectively. All calculations were carried out with an energy mesh cutoff of 300 Ry and a k-point mesh of $8 \times 6 \times 1$ in the Monkhorst–Pack scheme [36]. For the modeling we used a rectangular supercell of 198 C atoms. Calculations of formation energy were carried out by using a standard formula, $E_{form} = (E_{host+2nX} - (E_{host} + NE_{X2}))/n$, where $E_{host+2nX}$ is the total energy of the system after adsorption of 2n adatoms of species X, $E_{hos}$ is the total energy of the system before adsorption, and $E_{X2}$ is the total energy of an $X_2$ molecule in an empty box. We have also checked the contribution to the formation energy from zero-point energy correction and entropy. For the case of fluorination the sum of both of these corrections is less than 0.1 eV that is much smaller than enthalpy of formation and we did not take into account this correction in our further calculations. Calculations of migration barriers were performed by scanning of potential surfaces along migration pathway as it in details described for migration of various species on graphene in our previous work [33].

## 3. Fluorination versus hydrogenation

As the first step of our survey, we examined the energetics of the step-by-step fluorination and hydrogenation of graphene. As a single step of the process we take adsorption of a pair of adatoms, approaching from both sides of the graphene sheet, onto two neighboring carbon atoms (Fig. 1 inset). Previous calculations [6] have demonstrated that this configuration is the most energetically favorable for all monovalent species. In the case of hydrogen adsorption, adsorption of the first pair of hydrogen atoms is an endothermic process, in contrast with the very energetically favorable adsorption of the first pair of fluorine atoms (Fig. 2). This result is in accordance with the results of previous theoretical calculations [24,25]. Previous experimental results have also evidenced the rather easy fluorination of graphene [24] relative to hydrogenation [6]. For each subsequent step of functionalization we examined all possible positions of new adatom pairs and chose the configuration with the least total energy among them. The first step in the formation of these spot-like patterns corresponds with fluctuations in the curve of formation energy (Fig. 2), caused by the presence of distortions near the hydrogenated area on the graphene flat (Fig. 1). Further extension of the hydrogenated area is an exothermic process with consistent formation energy at each step. This result indicates the energetic unfavorability of coinciding hydrogenated and nonhydrogenated areas in a single graphene flat (see discussion in Ref. [36]). In the last steps of the hydrogenation process we also observe fluctuations in the formation energy curve arising from the hydrogenation of small nonhydrogenated islands; this situation is similar to that observed in the process of nanographene hydrogenation, which has similar energetics [36].

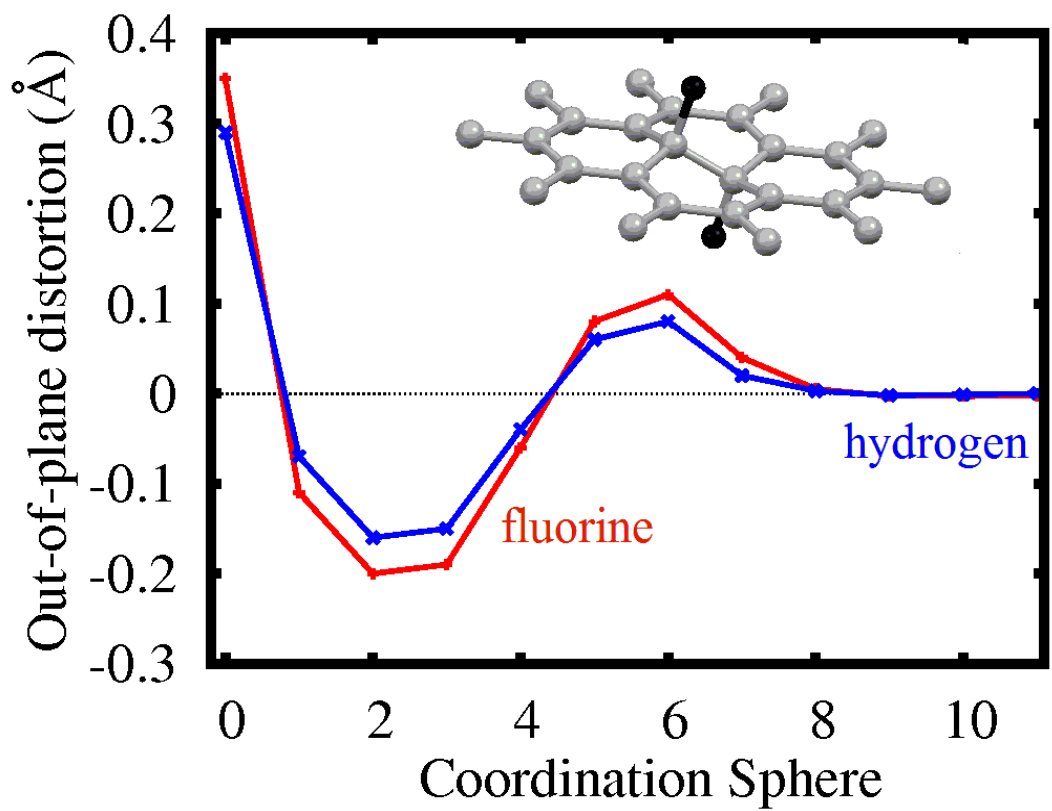

**Figure 1.** Graphene out-of-plane lattice distortions after chemisorption of a pair of fluorine or hydrogen adatoms versus coordination sphere. Inset: schematic illustration of an absorbed adatom pair.

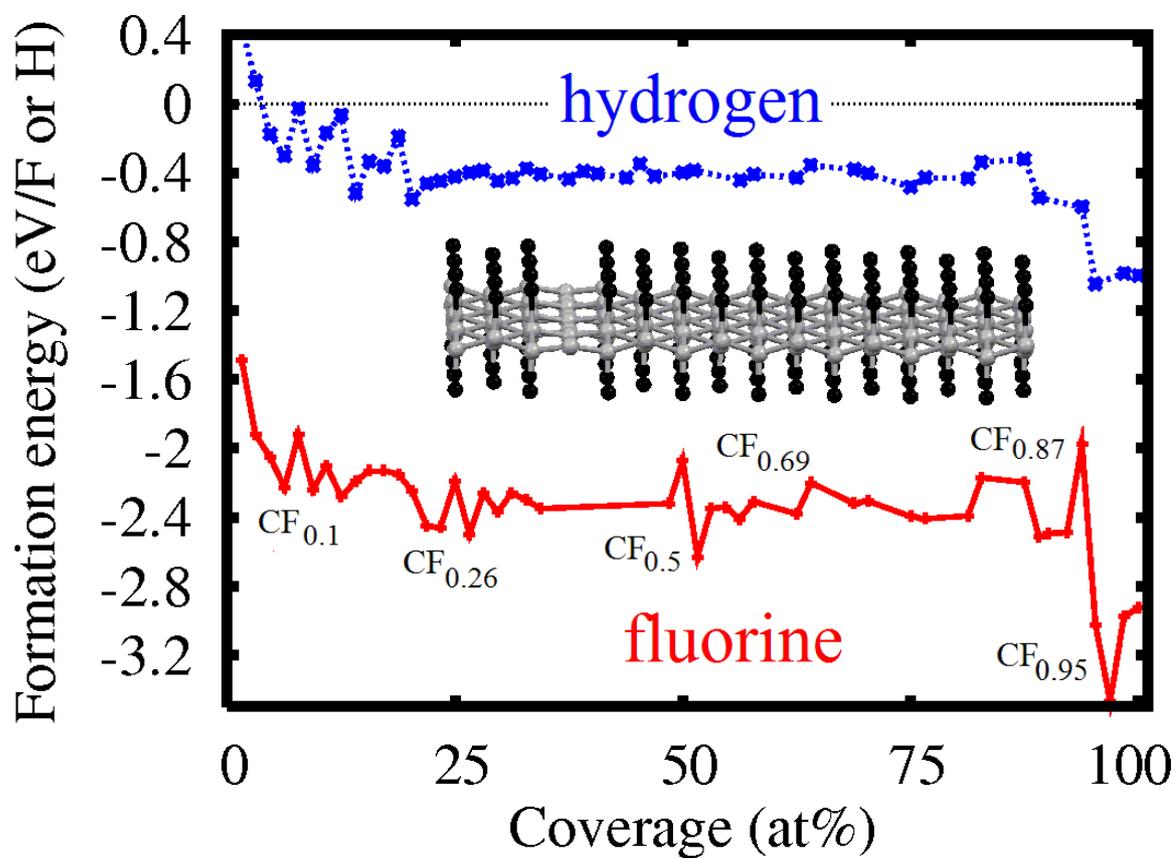

**Figure 2.** Step-by-step energetics of hydrogenation and fluorination of graphene. Inset: optimized atomic structure of a fluorographene with 93.5% fluorine coverage.

Adsorption of the pair of fluorine atoms leads to high-magnitude distortions of the graphene sheet (Fig. 1), which in combination with the very low formation energies of fluorine adsorption leads to completely different energetics of functionalization and atomic structure in partially fluorinated graphene. On one hand, the high-energy deformations of the graphene sheet makes further adsorption of fluorine atoms less favorable, but on the other hand, fluorine is a strong oxidative agent and covalent C–F bonds are much more stable than covalent C–H bonds. Step-by-step modeling of the fluorination process demonstrates the tendency of stripe-like patterns to form along the armchair directions (see Figs. 3, 4) during the earliest stages of fluorination. This strict anisotropy is caused by the greater difference between distortions in the zigzag and armchair directions in the case of fluorination, compared to that in the case of hydrogenation. Addition of the next pair of fluorine adatoms to complete wider ribbon-like patterns increases the formation energies because these patterns are rather stable, including uniform distortions and addition of new pairs of fluorine impurities in the vicinity of the edges of these patterns creates new distortions that are less uniform. As a result, higher formation energy is required during the first steps of the formation of a new line of fluorine adatoms along the edge of a complete ribbon. These results are in agreement with recent theoretical evaluation of the stability of some fluorine configuration on graphene [27]. The C/F ratios corresponding with the greatest magnifications of formation energy could be compared with the same values in experimentally obtained stable $CF_x$ structures. One of the largest increases of formation energy occurred in the case of increasing fluorine coverage above 50%, because the structure is rather stable when half the graphene sheet is fluorinated and half is not; this finding corresponded with experimental results showing that $C_2F$ structures were synthesized more easily and more frequently than other $CF_x$ configurations [16,17]. The increases in formation energy each time the relative uniformity of a fluorination pattern is disrupted by the addition of new members, acting in opposition to the exothermic character of the overall process, results in the observed multiplicity of $CF_x$ configurations. On one hand, fluorination could stop after the formation of any uniform pattern, but on the other hand, the absolute energetic favorability of fluorination allows the further expansion of ribbon-like fluorine patterns despite the relative increases in formation energy.

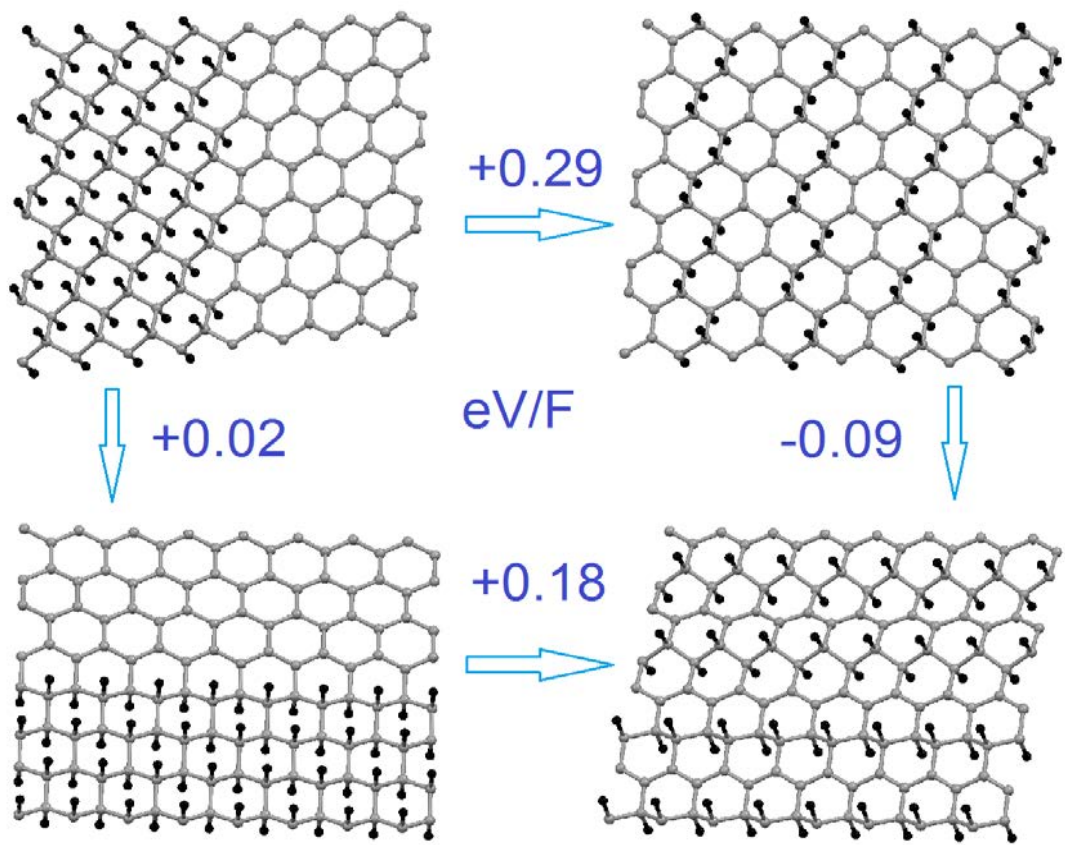

**Figure 3.** Energy differences (per fluorine atom) between various uniform configurations of $CF_{0.5}$.

One more experimentally observed feature of fluorination was the difficulty of obtaining 100% fluorination: C/F ratios no greater than about 10:9.5 could be obtained. Our calculations also demonstrate that the lowest formation energy corresponded to the $CF_{0.935}$ configuration, which represents the fluorination of almost the entire surface of graphene excepting one line of carbon atoms (Fig. 2 inset). In this configuration, all the carbon atoms in this line have one fluorinated carbon atom among their nearest neighbors and the atomic structure of the two other nearest neighbors is already greatly distorted; furthermore, the positions of all carbon atoms in this line are almost fixed by the environment that is fluorinated on both sides. Addition of the next pair of fluorine atoms thus would occur on these almost fixed carbon atoms, increasing the formation energy. The $CF_{0.95}$ configuration is about 0.5 eV more energetically favorable than the $CF_{0.935}$ configuration, whereas in the case of hydrogenation, the formation energies of the analogous configurations are nearly the same as each other. This phenomenon arises from the different C–C bond lengths (and thus different bond energies) in fluorographene and graphene. In the case of graphene these bonds are shorter (1.48 Å) than those in diamond (1.54 Å), which is used as an approximation of C–C distance in the case of $sp^3$ hybridization. In fluorographene, the C–C bonds (1.55 Å) are slightly longer than those in diamond. This situation is similar to that of graphene oxide: 100% coverage of graphene oxide by hydroxyl groups increases the C–C bond distance above that of diamond, increasing the formation energy and making incomplete oxidation energetically more favorable, as observed in samples of graphene oxide prepared by various methods [3]. Because the difference from the diamond C–C bond length is less for fluorographene than for graphene oxide, the maximal level of functionalization is closer to 100% for fluorographene.

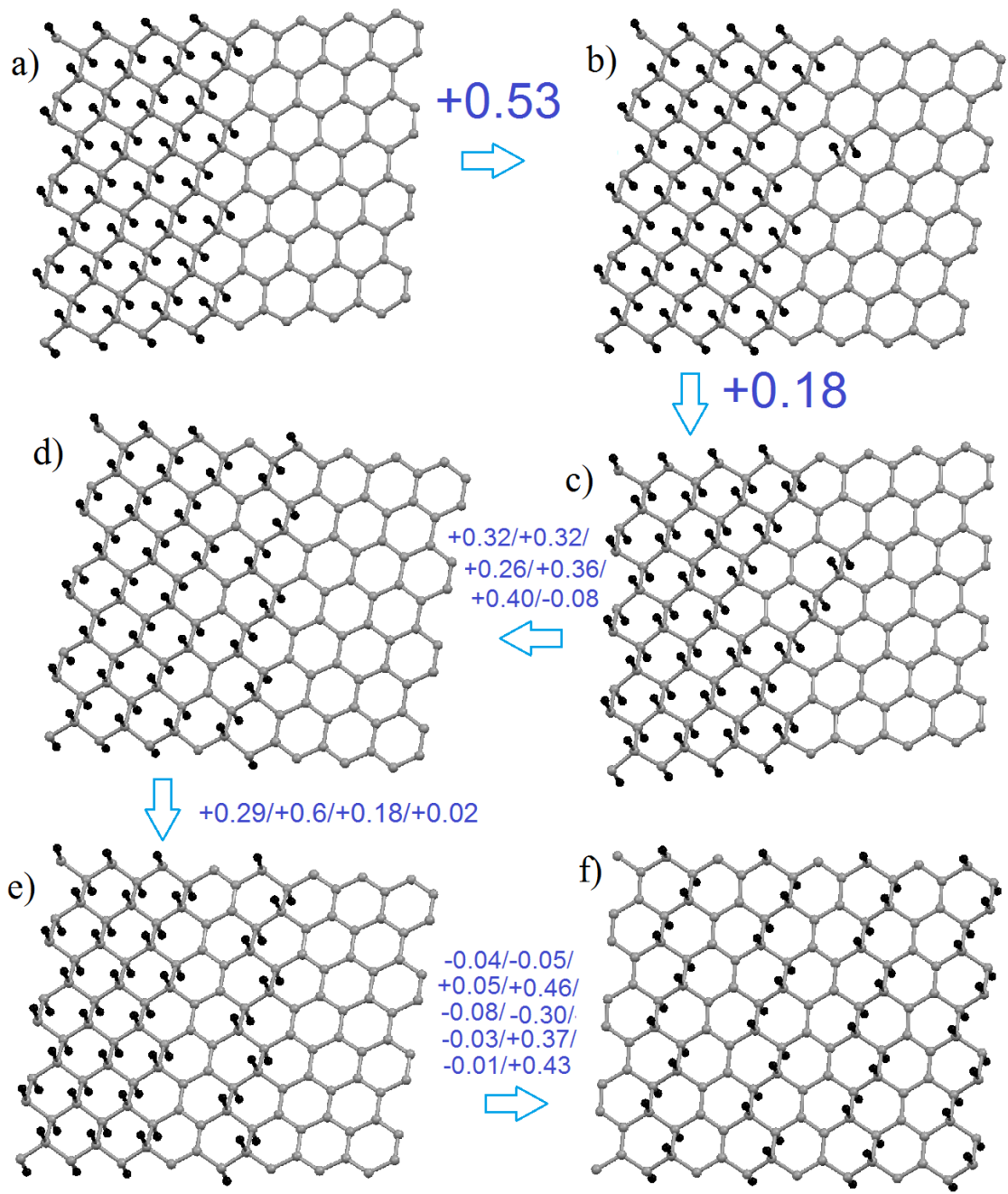

**Figure 4.** Energy costs of the transformation of armchair ribbon-like patterns of semifluorinated graphene (CF$_{0.5}$) to form single lines oriented along the armchair directions. Groups of multiple numbers represent the energies of omitted intermediate steps.

## 4. Stability of CF$_{0.5}$ configurations

The step-by-step modeling of fluorination demonstrates the energetic favorability of the formation of ribbon-like configurations, but experimental measurements [18] cannot evidence for semifluorinated graphene a preference between the two graphene axes, and also provides no information regarding the exact shapes of the uniform patterns (namely, of ribbon-like lines, or of alterations to the fluorinated and nonfluorinated lines; see Fig. 3). Previous experimental investigations [19-21] have suggested that ribbon-like patterns are likely to form, but have also demonstrated considerable amounts of fluorine adatoms (about 15% of fluorine content) assuming other configurations such as lines, pairs, or single atoms. These results encouraged us to examine other uniform configurations, namely ribbon-like and stripe-like patterns along both the armchair and zigzag directions. Our calculation results demonstrate that in the case of the orientation of ribbon-like patterns along the zigzag direction, the total energy increases only 0.02 eV per $C_2F$ unit relative to the armchair orientation (Fig. 3). The transformation of ribbon-like patterns to patterns of alternating fluorinated and nonfluorinated lines is less energetically favorable in both cases, but the energy differences among all configurations were rather low according to the modeling results. Note that the entropy of the systems almost does not change after rearrangement of fluorine patterns on graphene because vibrational degrees of freedom of fluorine adatoms remain the same and vibrational energies depends mainly from the strength of C-F covalent bonds.

In contrast to the phase transition in solids, in the present case transition from one uniform configuration to another is a step-by-step process. For example, consider the step-by-step process of separation of the outermost line of fluorine atoms from a ribbon-like pattern (Fig. 4). According to our calculation results, the first step of the transformation is the most energetically unfavorable because this step disrupts the united structure of the pattern (Fig. 4b); each of the further endothermic steps in this process (Fig. 4c,d) require less energy. Despite the positive sign of this step's energy, its magnitude is rather small, on the order of 0.5 eV. Further migration of this line of fluorine , separation by a second line from the ribbon-like pattern (Fig. 4e), and finally complete transformation of the ribbon-like pattern to a pattern of alternating fluorinated and nonfluorinated lines (Fig. 4f) each require energies below +0.5 eV if endothermic, and several steps of this process are exothermic. Estimations of the temperatures of chemical processes on a graphene substrate conducted in our recent work [38] demonstrate that the energy costs of about 0.5 eV obtained by means of DFT calculations for graphene correspond with processes able to occur at room temperature. Thus we can conclude that patterns formed along the armchair axis during fluorination are rather unstable at room temperature and could be partially decomposed despite their energetic favorability. However, deeper understanding of the stability of fluorinated patterns should also take into account the magnitudes of migration barriers and the effects of the fluorination extent.

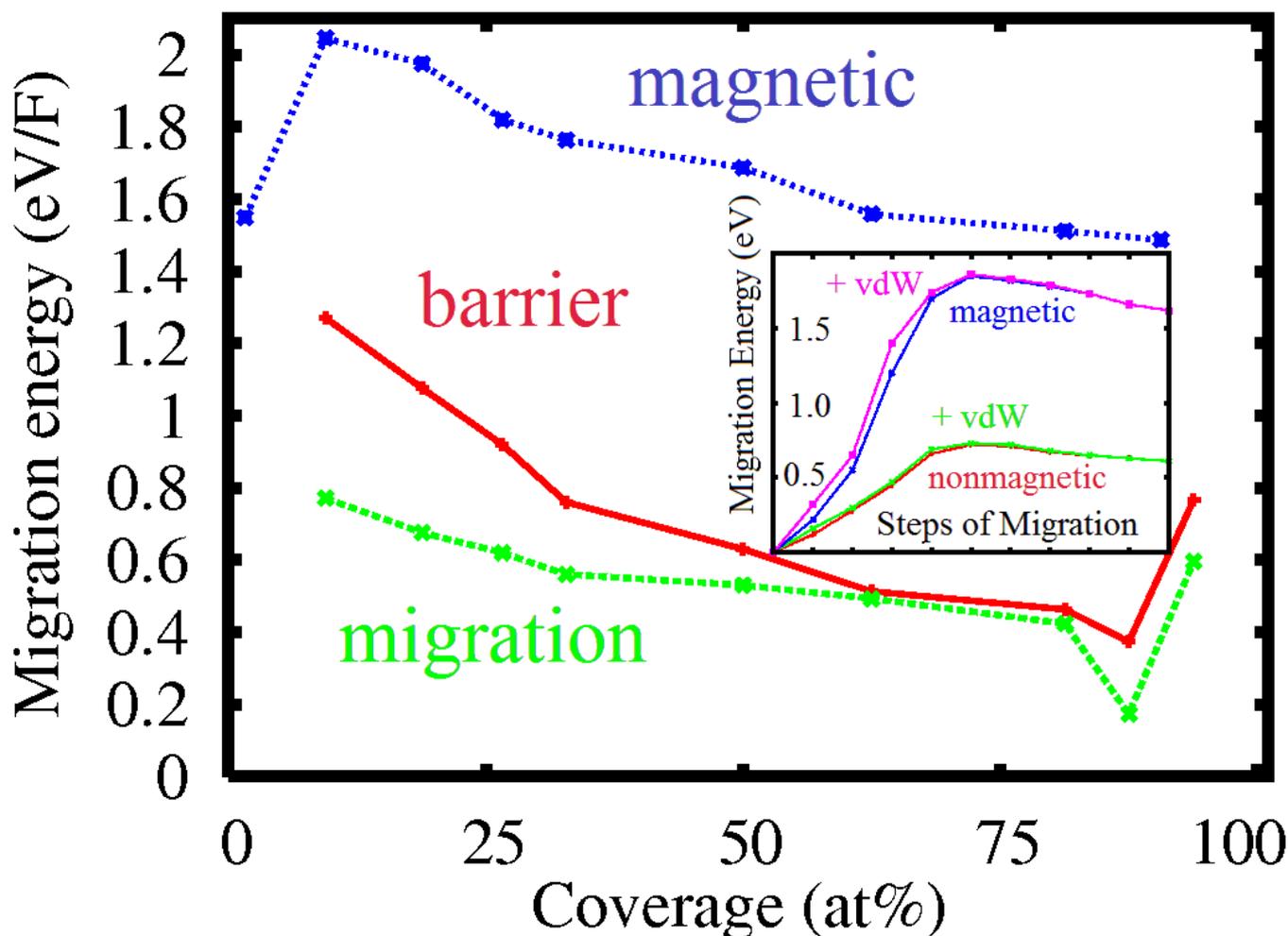

**Figure 5.** Energy costs and energy barrier to migration of a pair of fluorine atoms (see Fig. 1 inset) out of a fluorinated area, and energy cost of migration of a single atom from fluorinated areas with appearance of dangling bonds and magnetic moments, versus extent of fluorine coverage on graphene. On inset changes of migration energies along trajectory of migrations with and withoiut taking int account dispersion forces (+vdW).

## 5. Effects of fluorination extent on structural and magnetic properties of fluorographene

In any discussions regarding the possibility of structural changes, calculation of the energies of the initial and final configurations is not sufficient; calculation of the energy barriers to migration is also required [30]. We performed calculations of energy costs and migration barriers for movement of a pair of fluorine atoms away from a ribbon-like pattern (Fig. 4a,b) for all studied fluorination extents. The results of these calculations (Fig. 5a) demonstrated that both the energy cost and the migration barrier decrease with increasing fluorine content. The cause of this effect is the increasing distortion of the nonfluorinated area with increasing coverage. In cases of low coverage, a pair of fluorine atoms migrates from the narrow fluorinated area to the large and weakly disturbed nonfluorinated area, creating additional strong distortions there. Further fluorination corresponds with migration of the fluorine pair from a larger fluorinated area to a narrower and already distorted nonfluorinated area, decreasing the contribution to energy cost by means of the distortion of the graphene sheet. We have also checked the role of dispersion-forces [39] on migration barriers and find that it provides insignificant increasing of migration energies at initial stages and play almost no role for the heights energy barriers (see inset on Fig. 5).

The covalent chemical bond between carbon and fluorine is polar and also discussed as "semi-ionic" [25]. One of the results of fluorination is charge transfer from carbon to fluorine. The charge transferred from the carbon atom to fluorine is partially compensated by redistribution of the charge throughout the graphene membrane, including the nonfluorinated part. This wide distribution of charge is similar to the broad redistribution of the unpaired electron arising after adsorption of a single hydrogen atom [6]. Thus we can say that the nonfluorinated part of the partially fluorinated graphene is doped graphene, and the degree of doping changes significantly with changes in the extent of fluorination. The Mulliken populations of the carbon atoms in the nonfluorinated part of the graphene became closer and closer to those of the carbon atoms in the fluorinated part of graphene that make both parts similar that is provide decay of the barrier of migration from fluorinated to nonfluorinated part. One more feature to discuss is the drop of migration energies at the fluorination level of about 80–90%. This drop is similar to the decreasing of formation energies at these fluorination degrees and corresponding with favorability of migration of fluorine adatoms to remain two strongly distorted and strongly doped (about $-0.1$ e$^-$/C) lines of nonfluorinated carbon atoms. Further increase in the extent of fluorination leads to the formation of the robust $CF_{0.95}$ structure (see Fig. 2 inset) for which the approach of fluorine atoms over the carbon atoms in the single unsaturated chain is energetically unlikely.

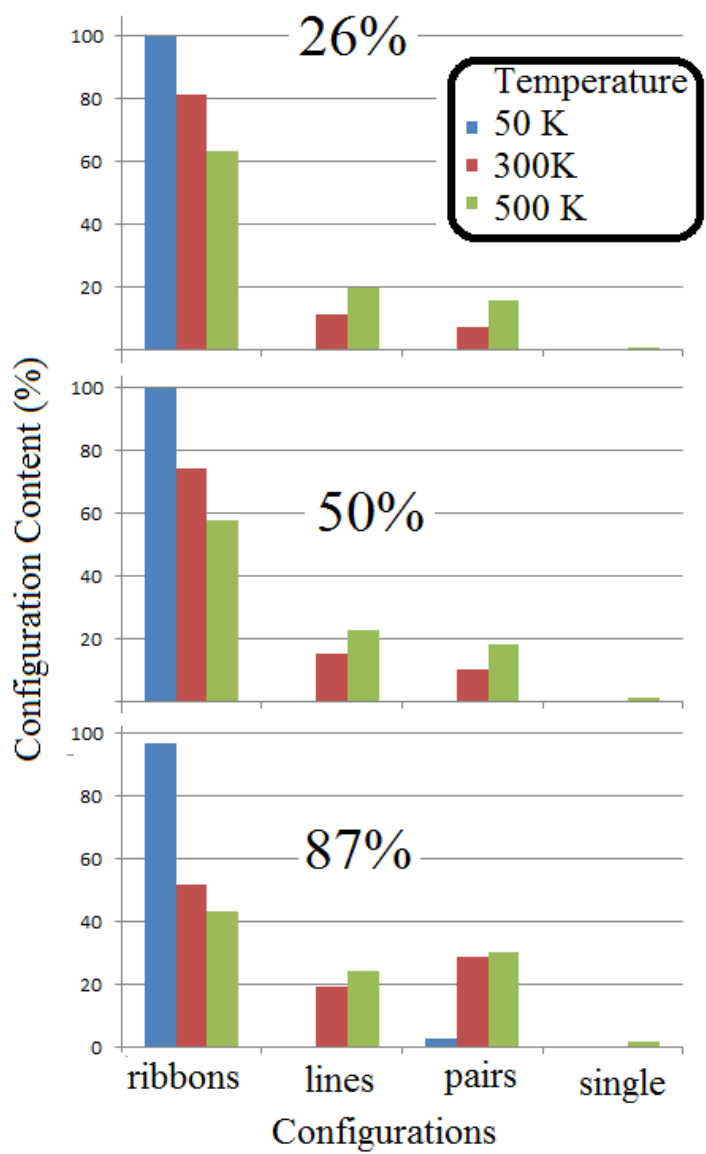

**Figure 6.** Boltzmann distribution of four main configurations of fluorine adatoms on graphene: ribbon-like and lines (see Fig.3), pairs on closest carbon atoms (see inset Fig. 1) and single adatons.

For evaluation of the effect of the temperature for mobility of fluorine adatoms over graphene we calculate Boltzmann distribution of configurations at 50, 300 and 500 K. We take into account four main types of configurations – ribbon like patterns and lines of fluorine adatoms (Fig. 3), pair of fluorine atoms on different sublattices (inset of Fig. 1) and single fluorine adatom. Results of calculations (Fig. 6) for $CF_{0.5}$ at room temperature are qualitatively close to results of experimental estimation of the ratio of different configurations in this compound. [20] Dependence of the stability of fluorine patterns from temperature could be used for practical application. At the lowest temperatures almost all fluorine atoms located within ribbon like patterns and at the temperatures above 200°C more than half of fluorine adatoms is within lines and pairs. Combination of the sensitivity to the change of temperature with thermal stability makes fluorinated graphene suitable for employment as temperature detector because changes of atomic structure provide changes of electrical and also magnetic properties. Temperature dependence of atomic structure should be also taken into account for discussion of fluographene as wear material and lubricate similarly to fluorinated graphite. Changes of mobility of fluorine adatoms on carbon flat can be also used for engineering of atomic structure of fluorinated graphene by fixation of obtained desirable structure by capping of fluorinated graphene by other layered materials of adsorption of molecules between fluorinated patterns. The last step of our work was to check the energetics of the magnetic configurations. In previous works two main mechanisms of intrinsic magnetism in graphene have been discussed — vacancies and the appearance of unpaired electrons on dangling bonds caused by the different numbers of species adsorbed on different sublattices [6]. As the simplest process yielding the formation of a magnetic configuration we describe the migration of a single fluorine atom from the ribbon-like pattern to the nearest carbon atom of the other sublattice. Because the numbers of fluorine adatoms on the different sublattices of graphene become unequal, a magnetic moment caused by the unsaturated dangling bonds appears. The results of our calculations (Fig. 5) demonstrate that for fluorine coverages above 10%, the energy cost of formation of a magnetic configuration decreases but remains considerably higher than the energy cost of migration of fluorine impurities to yield a nonmagnetic configuration. In other words, increasing the extent of fluorination facilitates the formation of magnetic configurations (see also Fig. 6), which explains the experimentally observed increases in magnetization of samples with increasing fluorine content [16]. Note that in contrast to the case of of transitional metals oxides where some methods of including of many-body corrections (such as DFT+U) [40] is feasible for obtaining correct magnetic structure in the case of light elements magnetism (also called *sp*-magnetism or *d0*-magnetism) more sophisticated methods such as used for obtaining of magnetic ground state of one-side fluorinated graphene is required. [41] Because mentioned method require to performing of some model calculations obtaining of realistic atomic structure of fluorinated graphene is the first step in theoretical description of its magnetic properties.

## 6. Conclusions

Based on the results of first-principles calculations we have discussed the difference between the fluorination and hydrogenation processes, finding that in the case of fluorination, the larger-magnitude out-of-plane distortions of the graphene sheet significantly inhibit the process of functionalization. In contrast with the formation of spot-like patterns during hydrogenation, fluorination proceeds by means of the formation of ribbon-like patterns along the armchair directions. Addition of further pairs to these metastable ribbon-like patterns distorts them and thus decreases the energetic favorability of initiating additional rows along their edges. This result explains the existence of various $CF_x$ structures that were easily formed in experiments. The combination of strong distortion of the nonfluorinated area and the doping of graphene caused by the polar nature of the C–F bonds decreases the energy cost of migration of fluorine atoms from stable patterns with increasing fluorination extent. The energy cost of formation of magnetic configurations also decreases with increasing fluorination extent. These results explain the existence of various configurations of fluorine adatoms in semifluorinated graphene and the increasing magnetization of fluorinated graphene with increasing fluorination extent.

**Acknowledgements** The work is supported by the Ministry of Education and Science of the Russian Federation, Project N 16.1751.2014/K